\documentclass[12pt,preprint]{revtex4-1} 

\usepackage{amsmath}  % needed for \tfrac, \bmatrix, etc.
\usepackage{amsfonts} % needed for bold Greek, Fraktur, and blackboard bold
\usepackage{graphicx} % needed for figures

\begin{document}

% Be sure to use the \title, \author, \affiliation, and \abstract macros
% to format your title page.  Don't use lower-level macros to  manually
% adjust the fonts and centering.

\title{Einstein's first gravitational field equation 101 years latter}
% In a long title you can use \\ to force a line break at a certain location.

\author{Juan Betancort Rijo}
\email{jbetanco@iac.es} % optional

\altaffiliation[permanent address: ]{} % optional second address
% If there were a second author at the same address, we would put another 
% \author{} statement here.  Don't combine multiple authors in a single
% \author statement.
\affiliation{Instituto de Astrof\'{i}sica de Canarias, E-38200 La Laguna, Tenerife, Spain.\\
Departamento de Astrof\'{i}sica, Universidad de La Laguna, E-38205 La Laguna, Tenerife, Spain.}

\author{Felipe Jim\'{e}nez Ibarra}
\email{fji@cca.iac.es} % optional
\altaffiliation[permanent address: ]{} % optional second address
% If there were a second author at the same address, we would put another 
% \author{} statement here.  Don't combine multiple authors in a single
% \author statement.
\affiliation{Universidad de La Laguna, E-38205 La Laguna, Tenerife, Spain.}

% See the REVTeX documentation for more examples of author and affiliation lists.

\date{\today}

\begin{abstract}

We review and strengthen the arguments given by Einstein to derive his first gravitational
field equation for static fields and show that, although it was ultimately rejected, it
follows from General Relativity (GR) for negligible pressure. Using this equation and
considerations folowing directly from the equivalence principle (EP) , we show how
Schwarzschild metric and other vacum metrics can be obtained immediately. With this results
and some basic principles, we obtain the metric in the general spherically symmetric case
and the corresponding hydrostatic equilibrium equation. For this metrics we obtain the
motion equations in a simple and exact manner that clearly shows the three sources of
difference (implied by various aspects of the EP) with respect to the Newtonian case and use
them to study the classical tests of GR. We comment on the origin of the problems of
Einstein first theory of gravity and discuss how , by removing it the theory could be made
consistent and extended to include rotations ,we also comments on various conceptual issues
of GR as the origin of the gravitational effect of pressure.

\end{abstract}

\maketitle % title page is now complete
\section{Introduction}

In the development of General Relativity (GR) Einstein (1907\cite{A.E.1907}) started  by
using the equivalence principle (EP) to analyse the ``apparent'' gravitational field in an
uniformly accelerated system.  In this analysis he found that the synchronized time
(coordinate time) could not agree with the local proper time, a fact from which it followed
immediately the gravitational redshift, the bending of light trajectories and the dependence
of light velocity on position when measured using coordinate time. In 1911\cite{A.E.1911} he
reviewed and expanded these ideas and conjectured that the effects found in the ``apparent''
gravitational field of an accelerated system, should also be present in a genuine
gravitational field as that of the earth or the sun. In particular, he predicted a
deflection of light rays grazing the sun of roughly one arc second. However, after a lengthy
and contorted path he will come to know that, unlike the gravitational redshift and the
dependence of the velocity of light on position, which hold exactly in a general static
field, the deflection of light ray had a hidden assumption on the flatness of space.
Depending on the curvature of space this deflection could be zero (conformally flat
space-time, corresponding to a negative spatial curvature) or twice the quoted value (full
GR, corresponding to positive spatial curvature around the sun). In 1912 \cite{A.E.1912I} he
assumed that a general static gravitational field could be described by a single scalar
field, namely, the position dependent velocity of light, $c(\vec{x})$, that he has found to
be the case for the field in an accelerated system. Again, the implicit assumption here was
that space remained flat and, as we shall discuss latter, this was bound to give rise to
difficulties. Einstein obtained the equation satisfied by the field $c(\vec{x})$ in an
accelerated system and through some considerations extended it to the case of a general
static field. The equation was:

\begin{equation} \nabla^{2}c(\vec{x})=4 \pi G c(\vec{x}) \rho(\vec{x}), \label{1Juan}
\end{equation} where $\rho(	\vec{x})$ is the matter density distribution.  The problem
was that one can not arrange (through suitable definition of mass and force within a
gravitational field) for the simultaneous conservation of energy and momentum. This
difficulties led Einstein to advance a modified equation fully consistent with
energy-momentum conservation, but not exactly consistent with the EP, which would be
satisfied only by sufficiently weak fields. Einstein considered the EP to be the happiest
thought of his live, thus having to forsake this principle was rather unpalatable to him. In
this state of inner dissatisfaction with his latest field equation, he started to muse on
the plausibility of his assumption concerning the Euclidean character of  space. Analysing
the intrinsic geometry of a rotating disk\cite{A.E.libro} he realized that, even in the
absence of a genuine gravitational field (i.e.  flat space-time), in the co-rotating system,
where an ``apparent'' gravitational field exist, the geometry of space can not be Euclidean.
The EP then led  him to believe that the same would be true for an actual gravitational
field. It seems that it was at this point that he fully realized that the EP implied the
tensorial character of the gravitational field, which is essential even in the case of
static fields. From this moment his research took a brisk turn. Once he realized that the
appropriate formalism for his problem was that of ``absolute calculus'' (the name given in
those day to de differential geometry) he set himself to command that calculus and initiated
a long an extremely arduous route culminating three and a half year latter ( in November
1915) with the fully-fledged theory of GR.  The story of these development is described in
some detail in books like Abraham Pais's\cite{A.Pais} Einsteins's scientific biography or
Howard $\&$ Stachel's\cite{Stachel} history of the development of GR  and it represented for
Einstein a novel style of research, not only with respect to GR, but with respect to all his
previous scientific career. Up to this time physical intuition played the guiding role; in
the future that role will be played mostly by  formal considerations. It was after GR was
obtained that the intuitions for the new physics was gained by extracting its
implications.However, the question of which portion of GR could have been anticipated by
simple consideration based on EP have, to our knowledge, not been fully clarified. More
precisely, the questions of the validity of Eq. (\ref{1Juan}) and whether the  derivation
Einstein gave for it was correct,  or which relevant results may be inferred from it, are
not found in the most widely used text on the conceptual development  of GR.

In this work we want to point out  that the Eq. (\ref{1Juan}) is correct (for negligible
pressure) and that the arguments given by Einstein to prove it from the EP are, with some
qualifications , correct, being some other assumption that Einstein took that led  to
contradictions. The goals that we pursue are three-fold: from the
conceptual point of view we want to explore how much can one learn about a static
gravitational field from simple (pre- Riemannian) considerations concerning the EP, as it
was the aim of the interrupted Einstein GR research program. From a technical point of view
we want to stress that Eq. (\ref{1Juan}) is a very useful one for static field, and that in
the case of spherical symmetry it leads immediately to rather interesting results. From the
historic point of view we want to remark that by a slight twist of events a theory of
gravitation accounting for most relevant facts (including the three classical test) would
have been available in 1912.

The work is organized as follows: In section 1 we review Einstein derivation of Eq.
(\ref{1Juan}) weighing and strengthening the arguments he gave and show that, in the case of
negligible pressure,  it follows from Einstein equations of GR. In section 2 we use Eq.
(\ref{1Juan})  together with a simple argument relating the $c$ field to the spatial metric
to derive de Schwarzschild metric. We also consider some other interesting results following
from Eq. (\ref{1Juan}). In section 3 we discuss Einstein's pre-Riemannian theories of
gravitation reviewing the arguments and the difficulties that he  found. We also consider
which could have been the course of development of GR if Einstein had been able to continue
his initial program, which involved considering particularly meaningful physical situation
of increasing complexity.

% AJP requires an abstract for all regular article submissions.
% Abstracts are optional for submissions to the "Notes and Discussions" section.

\section{Einstein's first gravitational field equation}

In a work published in February 1912\cite{A.E.1912I} Einstein obtained the gravitational
field which according to de EP must exist in an accelerated system with sufficient accuracy
to show that the Laplacian of the corresponding $c$-field must vanish. He did not elaborate
on the properties of that field beyond this result because he felt some uneasiness about the
fact that that gravitational field was not homogeneous. In a work by Born\cite{M.B.1909}
concerning the motion of rigid bodies in special relativity it have been shown that  in
order for a body to be moved rigidly, so that  every portion of it retain its size in the
comoving systems, different parts of the body  must experience different accelerations.  An
uniformly accelerated system must have a position dependent acceleration (although at a
given point it is constant in time) so that the gravitational field implied by de EP is not
homogeneous. Einstein knew this result in 1912 and its implication that the EP can only be
applied locally, but he was so  confused by it that he preferred to avoid mentioning it
explicitly. Instead, he just said that the field under consideration is of ``some specific
kind'', implicitly acknowledging that it is not an uniform field, as he had assumed
previously.

The structure of that field may be investigated using the simple methods used by Einstein,
but for a modern reader it may be easier to use Rindler \cite{Rindler}  metric for an
uniformly accelerated system. Using  Cartesian spatial coordinates and synchronized time
coincident with proper time at the origin we have:

\begin{eqnarray}
&d^{2}s& =g_{00}d^{2}t-d^{2}x-d^{2}y-d^{2}z;\nonumber \\ 
 &g_{00}&=c_{o}^{2}\left(  1+ 	\dfrac{a_{0}x}{c_{o}^{2}} \right)^{2}  \nonumber \\
 &c(x)& \equiv g_{00}^{1/2} ; \nonumber \\
&\vec{g}&=-\vec{\nabla}\phi= c_{0}^{2}\dfrac{\partial}{\partial \vec{x}} \ln g_{00}^{1/2}= - \dfrac{a_{0}}{1+a_{0}c_{0}^{-2}x}\hat{i}; \nonumber \\
&\phi &=c_{0}^{2}\ln g_{00}^{1/2}\nonumber \\
&c(x)&= c_{0}e^{\phi/c_{0}^{2}}.
\label{larga}
\end{eqnarray} 

where $\vec{g}$ is the gravitational field strength (i.e. the force on the unit mass at
rest) and $\phi$ the corresponding potential. We have considered an acceleration in the $x$
direction ( $\vec{i}$  being the unit vector in this direction). The relationship between
$\phi$  and  $g_{00}$ is a well-known one for static field (Landau-Lifshtz\cite{Landau}),
that could be derived, anyway, from simple considerations. $a_{0}$ is the acceleration at
the origin ($x=0$) and $c_{0}$ is the speed of light in vacuum measured with local proper
time (i.e. it is a constant). At $x$ the absolute value of the gravitational field strength
is $a_{0}(1+a_{0}c_{0}^{-2}x)^{-1}$, in particular at $x=-a_{0}^{-1}c_{0}^{2}$ the 
gravitational field goes to infinity and  is obvious that a solidly moving body can not
extend beyond that point. According to the EP, this result implies that the proper
acceleration (i.e. that measured in the comoving system) is  $a_{0}(1+a_{0}x)^{-1}$, that
depends on $x$. This is the result that stunned Einstein and that is confusing in at least
two respects: first it is the question of how is it possible that the different portions  of
the system accelerate at different rates. At any one time an inertial system (with
coordinate time given by local proper time synchronized as usual in special relativity) may
be found where all portions of the accelerating system are instantaneously at rest. This
have been assumed by Einstein in his treatment of the accelerating system, and it seems to
follow necessarily from the very definition of a bodily accelerated system. It is actually
true, but it is in apparent  contradiction with different part of the system having
different accelerations. Secondly, it is  puzzling that different parts of a system can accelerate at different
rates relatively to the instantaneously comoving inertial system without deformations. The
answer to the last question is as follows: assume that two portions of the accelerated
system, the one at $x=0$ and the one at $x=l$, accelerate at the same rate. If at some time
both point are at rest with respect  to an inertial system, at time $t$ each of these
portion would have displaced with respect to this system by exactly the same amount , since
both portions experience the same hyperbolic motion (constant proper acceleration).So the
distance between both portions as judged from the inertial system would always be  $l$. But
as $t$ increases the velocity of those portions with respect to the inertial system
increases and if in the inertial system the distance between the two portions remains equal
to $l$, this means, due to Lorentz-Fitzgerald contraction, that in the accelerating system,
or in the inertial system where the accelerating system is instantaneously at rest, the
distance between those portion must increase. We then see that in order for the system to be
``rigidly '' accelerated, so that the distance between both portion remains constant in it,
the trailing portion must experience a somewhat larger acceleration than the leading one, so
that as both portion gain speed with respect to an inertial system the distance between them
as measured in this system show the contraction corresponding to an intrinsically constant
size.The first  questions may be answered as follows:if the system is initially at rest with
respect to certain inertial one and at some time start to accelerate and after some time has
gained certain speed, it is possible to find another inertial system in which the
accelerating system is instantaneously at rest because, although the trailing portion of the
accelerating system accelerate at a  larger rate, this portion started its motion latter
than the leading one as judged from this latter inertial system, due to the difference in
simultaneity with respect to the former inertial system.

It  is convenient to discuss these questions because they are interesting in themself and
because it help understanding Einstein reluctance to discuss the global properties of the
gravitational field under consideration.

What he did to avoid these difficulties was to considerer the local transformation from the
comoving inertial to the accelerating system around $t=0$, $x=0$ at second order in $t$. He
found for small values of $x$ (i.e. $x<<c_{0}^{2}a^{-1}$)

\begin{equation*}
c \simeq c_{0}+\dfrac{a}{c_{0}}x
\end{equation*}

$a$ is, in fact, the proper acceleration at $x=0$ (what we called $a_{0}$ in Eq.
(\ref{larga})), but Einstein avoided the question of the dependence of $a$ on $x$. From 
this result he concluded that  for this field the Laplacian of $c$ vanishes. This conclusion
seems rigorous, because the same arguments can be given for any value of x. In fact, the
exact expression for $c$, given in Eq. (\ref{larga}), has vanishing Laplacian.

Having studied the gravitational field in an accelerated system he went on to consider the
general static case. To this end he made the crucial assumption that  the geometry of space
was Euclidean so that the situation could be described with just one function, namely, the
c-field. It is interesting to note that although this preliminary theory is sometime
referred to as ``scalar'', $c(x)$ it is in fact a component of a second rank tensor in four
dimension ($c(x)=g_{00}^{1/2}(x)$) and, although Einstein did not possess this language at
the time, he was well aware of the fact that $c(x)$ it is an scalar only with respect to
purely spatial transformation. For a general gravitational field Einstein assumed that, as
for the accelerated system, the Laplacian of $c(x)$ vanished outside the matter
distribution. This may seem a big jump, and Einstein did not comment on its plausibility,
but it may be supported by the following considerations: in an accelerated system the
Laplacian of $c$ is equal to zero even for arbitrarily large fields (as $x$ goes to
$-a_{o}/c^{2}_{0}$ the strength of the field goes to infinity). For a genuine gravitational
field, we know that in the weak field limit ($\phi/c^{2}_{0}\ll 1$) Newtonian gravitation
hold, so that we have:

\begin{equation*}
\nabla ^{2} \phi \simeq 0;\quad  \nabla ^{2}c \simeq 0
\end{equation*}

By transforming the weak field generated by a distribution of masses to an accelerated
system it could be shown that the resulting field, which is non trivial and non-weak also
satisfies that $\nabla^{2}c=0$.

 From a formal point of view one could say that apart from $c$ itself the only other scalar
(spatial) that can be formed with $c$ is the modulus of its gradient, but the presence of
any of this two terms on the field equation in vacuum would be in contradiction with the EP,
since we would not have he correct field equation in the case of an accelerating system. We
can therefore say that, if not rigorously provable, the equation $\nabla^{2}c=0$ for vacuum 
is at least very highly plausible.

In the presence of matter it is clear that in the weak field limit we must have:

\begin{equation}
c \simeq c_{0}(1+\dfrac{\phi_{0}}{c^{2}_{0}});\quad \nabla^{2}c\simeq\dfrac{4 \pi G \rho}{c_{0}}.
\label{5Juan}
\end{equation}

where $\rho$ stands for energy density (all through the rest of this work). For the general
static case Einstein proposed the equation:

 \begin{equation}
\nabla^{2}c=k c \rho, \quad \textrm{where} \ k=\dfrac{4 \pi G}{c_{0}^{2}}.
\label{6Juan}
\end{equation}

 The  derivation he provided for it is , such as it stands (at least in the English
 version\cite{A.E.1912I}) incorrect. Einstein asserted that  both sides of Eq. (\ref{6Juan})
 should be homogeneous in $c$ because ``it follows immediately from the meaning of $c$ that
 $c$ is determined only up to a constant factor that depends on the constitution of the
 clock with which one measures $t$ at the origin of $K $ (the reference system)''. What
 strictly follows from this observation is merely the trivial fact that both sides of Eq.
 (\ref{6Juan}) must have the same dimensions so that the equation hold for any choice of the
 time unit. However,  Eq. (\ref{6Juan}) may be inferred correctly from the fact that any
 point can be taken as origin of the system so that at that point coordinate time agrees
 with proper time and, therefore, $c=c_{0}$. The function $c$ is defined up to a constant
 factor not because time units can be chosen arbitrarily (which it is obviously true, but
 lack physical implications) as the quotation say, but because the origin of the system can
 be chosen arbitrarily (which lastly depend on the E P), as the text most probably was
 intended to mean. In fact, interpreting ``the constitution of clock" as concerning not the
 physical clock at the origin, but the procedure used to stablish the time coordinate (that
 involves choosing an origin), this would just be the meaning of the quotation. Here we see
 a simple instance of the ubiquous and confusing statement that general covariance is an
 important component of GR. In fact, the physical content of this theory lays entirely on
 the EP in its strong version, that concerns also the gravity field equations and not merely
 the matter motion (weak version). General covariance is a  property of the formalism and
 lack physical meaning. What is physically meaningful is the fact that that general
 covariance is displayed by expressions containing only the metric tensor and quantities
 describing matter. The meaningful fact is that  GR equations do not contain elements (i.e.
 fields) not associated with matter or gravity that  could imply that the equations for
 matter and gravity did not have the same form (locally) in all freely falling systems. In
 other words, the content of ``general covariance'' is simply the absence of foreign
 elements in the equations for matter and gravity, which is a fact implied by EP (strong).

 In the presents case, as we have said, the relevant fact to prove that Eq. (\ref{6Juan})
must have the stated form (for negligible pressure) it is not the conventional one that this
equation must hold for any choice of units, but the fact, implied by the EP (strong), that
$c$  field can be obtain using a coordinate time that may be chosen to agree with the proper
time at any point. 

 In summary, we have seen that the arguments given by Einstein to obtain Eq. (\ref{6Juan})
are  rather cogent, which is not the case for the arguments that led him to drop this
equation, as we shall latter see.

 The arguments leading to Eq. (\ref{6Juan}) are strong enough  to leave almost no doubt
about its validity. Now we shall see that this equation followes from the equations of GR
(Einstein`s equations) when pressure is negligible. These equations can be written in the
form\cite{Landau}

\begin{equation}
R_{ij}=\dfrac{8 \pi G}{c_{0}^{4}} \left(  T_{ij} - \dfrac{1}{2} g_{ij} T \right) 
\label{7Juan}
\end{equation}

where $R_{ij}$ is the Ricci Tensor (i, j run from 0 to 3 ), $T_{ij}$ is the energy-momentum
tensor, and T is its trace ($ T\equiv T^{i}_{i} $). Using now the following expression for
the Riemann tensor\cite{Landau}:

\begin{equation}
R_{iklm}= \dfrac{1}{2} \left(  \dfrac{\partial ^{2}g_{lm}}{\partial x^{k}x^{l}}+
\dfrac{\partial ^{2}g_{kl}}{\partial x^{i}x^{m}} -
\dfrac{\partial ^{2}g_{il}}{\partial x^{k}x^{m}} -
\dfrac{\partial ^{2}g_{km}}{\partial x^{i}x^{l}} \right) + g_{np} \left( \Gamma_{kl}^{n}\Gamma_{im}^{p}-\Gamma_{km}^{n}\Gamma_{il}^{p}\right)
\label{8Juan} 
\end{equation}

and choosing locally Cartesian coordinates at any given point  ($g_{\mu\nu}=-\delta_{\mu \nu}$;
$\dfrac{\partial g_{\mu \nu}}{\partial x^{\lambda}}=0$, with $\mu, \nu , \lambda$ from 1 to
3) we have for $R_{00}$

\begin{eqnarray}
R_{00}=g^{il}R_{i0l0}=-\dfrac{1}{2} g^{il} \left( \dfrac{\partial ^{2}g_{00}}{\partial x_{i}x_{l}}\right)+g_{np}g^{il} \left(\Gamma^{n}_{0l}\Gamma^{p}_{i0}-\Gamma^{n}_{00}\Gamma^{p}_{il}\right)=\nonumber \\ 
=-\dfrac{1}{2}g^{ii}\dfrac{\partial^{2}g_{00}}{\partial x^{2}_{i}}+g_{nn}g^{ii} \left(\Gamma^{n}_{0i}\Gamma^{p}_{i0}-\Gamma^{n}_{00}\Gamma^{p}_{ii}\right)
\label{9Juan} 
\end{eqnarray}

where we have used the fact that $g_{ij}=0$  for  $i\neq j$. From the definition of
$\Gamma$'s we have:

\begin{equation*}
\Gamma^{\mu}_{0 \nu}=\Gamma^{0}_{\mu \nu}=0;\quad \Gamma^{0}_{\mu 0}=\Gamma^{0}_{0 \mu}= \dfrac{1}{2}g^{00}\dfrac{\partial g_{00}}{\partial x^{\mu}} 
\end{equation*}

where use has been made of the static character of the metric (i.e. $\dfrac{\partial
g_{ij}}{\partial x^{0}}=0$). We then have for $R_{00}$:

\begin{equation}
R_{00}=-\dfrac{1}{2}g^{\mu \mu }\dfrac{\partial^{2}g_{00}}{\partial x^{2}_{\mu}}+g^{\mu \mu}g_{00} \left(\Gamma^{0}_{\mu 0}\right)^{2}.
\label{Juan10}
\end{equation}

But in the locally Cartesian spatial coordinates that we are using the first term is simply
one half of the Laplacian of $g_{00}$ and the second:

\begin{equation*}
g^{\mu \mu}g_{00} \left(\Gamma^{0}_{\mu 0}\right)^{2}=-\dfrac{1}{4g_{00}}\left(\dfrac{\partial g_{00}}{\mid g_{\mu \mu } \mid^{1/2}\partial x^{\mu}}\right) ^{2}.
\end{equation*}

where we have use the fact that $g^{\mu \mu}=\left(g_{\mu \mu} \right)^{-1}$ (note that this
expression is valid for any orthogonal system), is simply $-1/(4g_{00})$ times the square of
the gradient (this involves a definition of gradient which is not the most widely used, but
it has the most direct meaning and is the one used through this work). In the usual
three-dimensional notation:

\begin{equation}
R_{00}=\nabla ^{2}g_{00}- \dfrac{1}{4g_{00}}\vert \vec{\nabla}g_{00}\vert ^{2}
\label{11Juan}
\end{equation}

Using the relationship:

\begin{equation*}
\nabla ^{2}f(\phi)=\dfrac{df}{d\phi}\nabla ^{2}\phi +\dfrac{d^{2}f}{d\phi^{2}}\vert\vec{\nabla 	\phi}\vert ^ {2}.
\end{equation*}

We finally obtain for $R_{00}$

\begin{equation}
R_{00}=g_{00}^{1/2}\nabla ^{2}g^{1/2}_{00}.
\label{12Juan}
\end{equation}

We have carried out this derivation in a locally Cartesian system, but, since $R_{00}$  is a
scalar for spatial transformations, it is clear that Eq. (\ref{12Juan}) must hold in any
system of spatial coordinates.

Assuming that the source is a perfect fluid with energy density $\rho$ and pressure $P$, we
have for $T$:

\begin{equation*}
T=\left(\rho - 3P\right).
\end{equation*}

If the fluid was not perfect we would only need to change $P$ by its average value in the
three principal direction. We then have:

\begin{equation*}
T_{00}-\dfrac{1}{2}g_{00}T=g_{00} \rho -\dfrac{1}{2}g_{00} \left(\rho -3P \right)=\dfrac{g_{00}}{2}\left(\rho +3P \right).
\end{equation*}

Inserting this and de Eq. (\ref{12Juan}) into Eq. (\ref{7Juan}), we find:

\begin{equation}
\nabla ^{2} g_{00}^{1/2}=\dfrac{4\pi G}{c_{0}^{4}} g^{1/2}_{00}\left(\rho +3P \right)
\label{13Juan}
\end{equation}

Noting that Einstein $c$ field is just $g_{00}^{1/2}$, we see that in the case of negligible
pressure this equation reduces to Eq. (\ref{6Juan}), that had been obtained by Einstein from
simple considerations. The pressure term is related to the delicate issue of the
gravitational effect of pressure. We shall latter comment on its origin in GR and on the
fact that it doesn't seem possible to derive it from simple consideration concerning static
fields.

It is interesting to note that if a $\lambda$ term was considered, that is, if we added to
the left hand side of Eq. (\ref{7Juan}) a term of the form $\lambda g_{ij}$ ($\lambda$ being
a constant), in Eq. (\ref{13Juan}) we should have $\rho - \dfrac{\lambda c_{0}^{2}}{4 \pi
G}$ in place of $\rho$.

We may summarize this section by saying that the arguments given by Einstein in
1912\cite{A.E.1912I} to derive  Eq. (\ref{6Juan}) were quite solid and that, as a matter of
fact, this equation is valid in GR when pressure is negligible, as it is apparent by
comparing Eq. (\ref{6Juan}) with the exact equation for $g_{00}$ in an static field in GR,
given by Eq. (\ref{13Juan}). This equation is quite meaningfull both because it is
instrumental in the explanation of several basics facts, as we shall show, and because of
its close relationship with an equation that played a role in the development of GR (Eq.
(\ref{6Juan})).The fact that this equation appear neither on chapters on statics fields in
the best known textbook, nor in the books known to us on the historic and conceptual
development of GR is quite surprising. This is still more difficult to understand noting
that it is a well-known result\cite{S.Ron}  that Eq.(\ref{12Juan}) holds for a metric where
$g_{00}$ it is the only non trivial coefficients. As we have seen, this could have been
easily generalized to any static metric and with it obtain Eq. (\ref{13Juan}).

\section{Spherically symmetric field: derivation of Schwarszchild metric}

Shortly after obtaining Eq. (\ref{6Juan}) and then rejecting it, Einstein started
questioning the assumption that in a gravitational field space remains flat. After some 
groping in the dark he came to the conclusion that the spatial metric must be integrated
into a common structure with the ``proper'' gravitational field, $g_{00}$. This structure is
the four dimensional tensor $g_{ij}$, that in a sense represent the gravitational field in
GR. Therefore, in this theory there are ten different ``gravitational potentials'', namely,
the ten algebraically independent component of a symmetric second rank tensor in four
dimension, and to determine them ten equations are needed. This can not be achieved if the
field source was only the matter density; a tensorial source is needed. It is then obvious
that this source must be the energy momentum tensor, $T_{ij}$ and the ten needed equation
are those given in Eq. (\ref{7Juan}).

The ``proper'' gravitational potential is still given by $g_{00}$ (more precisely
$c_{0}^{2}/2 \ln g_{00}$), in the sense that the gravitational field strength,$\vec{g}$, is
given by minus its gradient. But this quantity only give us the force that must be exerted
upon the unit of mass to keep it at a fixed position within the field, while all $g_{ij}$
are needed to obtain particles trajectories. Furthermore, even if we were only interested in
$g_{00}$, to obtain it by integrating Eq. (\ref{6Juan})(or rather, its exact counterpart Eq.
(\ref{13Juan})) the spatial metric coefficients, $g_{\mu \nu}$, are needed, since they enter
in the explicit expression for the Laplacian. Thus, Eq.(\ref{6Juan}), although it is exact
(for negligible pressure) in any static field it is not of much use without integrating it
simultaneously with all the other equation entering in Eq. (\ref{7Juan}).

For spherically symmetric fields, however, we shall show that Eq. (\ref{6Juan}) determines
uniquely the whole metric (with a simple additional assumption in the non-vacuum case).
Choosing appropriate polar coordinates, a metric with spherical symmetry can be reduced to
the form:

\begin{equation}
d^{2}s=g_{00} d^{2}t-g_{rr}d^{2}r-r^{2} \left( d^{2}\theta + \sin^{2} d^{2}\phi \right)
\label{14Juan}
\end{equation}

where $\theta , \phi$ are the usual polar angles and $r$ is a radial coordinate defined so
that the  area of a sphere of radius $r$ is $4 \pi r^{2}$. There are only two undetermined
metric coefficients, $g_{00}$ and $g_{rr}$, the others being predetermined by symmetry
and the choice of coordinate. 

If we could find a relationship between this two coefficients, we could use Eq.
(\ref{13Juan}) to completely determine the metric. We shall now see how to obtain a
relationship between $g_{00}$ and $g_{rr}$ in the vacuum case (Schwarzschild metric) by
means of the following argument:

Consider the flat time-space (i.e. Minkowskian) and the usual Galilean reference system,
$K$, within it. Now consider another system, $K^{\prime}$, that can be described as a
continuous set of locally Galilean systems,  $k^{\prime}(\vec{x})$, so that at each point ,
$\vec{x}$, (Cartesian coordinates with respect to $K$), the local system
$k^{\prime}(\vec{x})$ at time $t_{0}$  is moving  with respect to $K$ radially from the
origin with speed $V(r)$ ($r\equiv \vert\vec{x} \vert$, $\vec{V}(\vec{x})$ is parallel to
$\vec{x}$).Let us study the spatial geometry in system $K^{\prime}$. In system $K$ space is
flat, so the length of a circumference with radius $r$ in, let us say, plane $z=0$ and
centred at the origin is simply $2 \pi r$. In system $K^{\prime}$ the length measurements
are carried out with rods comoving with the local Galilean system $k^{\prime}(\vec{x})$. To
measure the length of the circumference this rods must be set at point $\vec{x}$
perpendicularly to the velocity of $k^{\prime}(\vec{x})$ with respect to $K$, therefore,
rods measuring  the circumference have the same length in $K$ and in $K^{\prime}$ and,
consequently, the length of the circumference must also be $2\pi r$ in this last system.
However the radius of the circumference in $K^{\prime}$ is now longer than $r$, because in
measuring it the rods must be parallel to the velocity of $k^{\prime}(\vec{x})$ with respect
to $K$ so that in this system those rods are affected by the Lorenz-Fitzgerald contraction.
Therefore, more rods are necessary to cover the radius in $K^{\prime}$ than in $K$.More
precisely ,the proper radius, $\chi$, in $K^{\prime}$, due to the contraction is given by:

\begin{equation}
\chi (r)= \int^{r}_{0} \dfrac{dr^\prime}{\sqrt{1-\left(\dfrac{V(r^{\prime})}{c_{0}}\right)^{2}}}.
\label{15Juan}
\end{equation}

From the differential relationship between $\chi$ and $r$ we have immediately:

\begin{equation*}
g_{rr}(r)= \left[1-\left (\dfrac{V(r)}{c_{0}}\right)^{2} \right]^{-1/2}.
\end{equation*}

On the other hand, the relationship between proper time, $\tau$, and the time coordinate in
system $K$, $t$, is given by:

\begin{equation*}
d\tau \equiv \dfrac{g_{00}^{1/2}}{c_{0}} dt= \left[1-\left (\dfrac{V(r)}{c_{0}}\right)^{2} \right]^{1/2}dt.
\end{equation*}

So we have:

\begin{equation}
g_{rr}=c_{0}^{2}g_{00}^{-1}.
\label{16Juan}
\end{equation}

It must be noted that the time coordinate in system $K^\prime$ is also $t$, the Galilean
time in $K$. So, by the geometry  of the space in system $K^\prime$ we mean the metric of
hypersurface $t=t_{0}$. This metric is not constant, because as the local systems move away
from the origin the relationship between  $V$ and $r$ changes. However, to our purposes it
will suffice to obtain the metric at $t_{0}$; a metric that we will compare with the static
metric that we are going to discuss next.

 In the initial formulation of the EP the effect of a constant gravitational field are
consider equivalent to the non-inertial effects in an uniformly accelerated system. In this
system, although different parts of it are in some sense at rest with respect to each other,
when a particle, for example  a photon, move from A to B, by the time the photon arrives at
B the inertial system  in which B is instantaneously at rest is moving with respect to that
in which A was at rest at the emission time. This fact determines the metric in the
accelerated system ($g_{00}$ being the only non-trivial coefficient). In a similar manner,
one way  interpret the spherically symmetric metric in vacuum as the metric in $K^{\prime}$
at $t_{0}$, which is not constant, forced to become constant by the presence of the
gravitational field, which in the process turns space-time from Minkowskian  to a curved
one. It must be noted that for this interpretation to work $V(r)$ must be equal to the
velocity attained by a particle that start falling at infinity with zero velocity by the
time it reach radii $r$, although this is immaterial with respect to relationship
(\ref{16Juan}).

In  system $K^{\prime}$, although the metric is instantaneously equal at time $t_{0}$ to
that in the gravitational field, the redshift experienced by light when moving within the
system it is not equal to that for the gravitational field (not even for infinitesimal time
of flight), because this redshift depends not just on the metric in $K^{\prime}$ at $t_{0}$,
but  also on its evolution. For example, for the redshift experienced between $r$ (at
$t_{0}$) and infinity in $K^{\prime}$ we have (with $c_{0}=1$):

\begin{equation*}
1+\dfrac{\Delta \lambda}{\lambda}= \dfrac{1+V(r)}{\sqrt{1-V(r)^{2}}} \equiv \gamma(r) \left(1+V(r)\right)
\end{equation*}

where the factor $1+V(r)$ is the classical Doppler effect associated with the fact that in
$K^{\prime}$ the local systems are actually moving. The $\gamma$ factor correspond to the
transversal Doppler effect, associated with the difference between the time at the local
system and that at infinity, which corresponds to the redshift in the gravitational field.

It is possible to select a ``reference system'' , $K^{\prime \prime}$, in Minkowski space
whose metric is in closer relationship with that in a gravitational field. The system
$K^{\prime \prime}$ can be conceived as a discrete set of local systems of $K^{\prime}$ with
small but finite size that for a short period of time, $\Delta t$, move with velocity $V(r)$
or $-V(r)$ and then instantaneously change its velocity to  $-V(r)$ or  $V(r)$ . The metric
in this system and the redshift of light propagating between any two points averaged over
times larger than $\Delta t$ (the classical Doppler averages out) is the same as in the
gravitational field.

What follows from this interpretation is that in  vacuum the effect of the gravitational
field on space and time are both implied by a single quantity (interpreted here as $V(r)$)
and, therefore, $g_{rr}$ and $g_{00}$ are related, being Eq. (\ref{16Juan}) the necessary
form of this relationship. We shall see that this depends on the fact that in vacuum both
$g_{rr}$ and $g_{00}$ at $r$ depends on the enclosed mass, which is a constant. On the other
hand, in the general spherically symmetric case, $g_{rr}$ depends on the enclosed mass but
$g_{00}$ (normalized to the center of the cloud) depends on the  enclosed mass and
pressure-volume  (furthermore, even without pressure $g_{rr}$ depends just on $M(r)$(see Eq.
(\ref{Juan21})), while $g_{00}$ depends on $M(r^{\prime})$ for $r^{\prime} \leq r$) and no
general relationship between $g_{rr}$ and $g_{00}$ can exist, as we shall see by showing
that the contrary assumption leads to inconsistencies. In the vacuum case, however, Eq.
(\ref{16Juan})must  hold for any value of the cosmological constant and in the case of a
charged mass ``point'' (REF Reissner-Nordstron metric).

Note that for this relationship to be valid it is implicit that coordinate time have been
chosen so that $g_{00}$ goes to one at infinity.

Using Eq. (\ref{16Juan}) in Eq. (\ref{6Juan}), for vacuum (with $\lambda =0$), we may
immediately derive the whole metric Eq. (\ref{14Juan}), which has now been reduced to just
one independent quantity, $g_{00}$:

\begin{equation}
\nabla ^{2} c \equiv \nabla ^{2}g_{00}^{1/2}=\dfrac{1}{r^{2}g_{rr}^{1/2}} \dfrac{\partial}{\partial r} \left( \dfrac{r^{2}}{g_{rr}^{1/2}} \dfrac{\partial g_{00}^{1/2}}{\partial r}\right)=\dfrac{1}{r^{2}} \dfrac{g_{00}^{1/2}}{c_{0}^{2}} \dfrac{\partial}{\partial r}\left(r^{2}g_{00}^{1/2} \dfrac{\partial g_{00}^{1/2}}{\partial r}\right)=0.
\label{17Juan}
\end{equation}

where the first equality come from the  expression for Laplacian in spherical coordinates
for a quantity, $g_{00}^{1/2}$, that  is spherically symmetric, and the last comes from
using Eq. (4).

A first integration of Eq. (\ref{17Juan}) give:

\begin{equation*}
r^{2}g_{00}^{1/2}\dfrac{\partial g_{00}^{1/2}}{\partial r}= r^{2}\dfrac{1}{2}\dfrac{\partial g_{00}}{\partial r}=A
\end{equation*}
where $A$ is a constant. Integrating again:
\begin{equation*}
g_{00}=-\dfrac{2A}{r} +B.
\end{equation*}

With the usual convention of taking the coordinate time to agree with proper time at
infinity (which is necessary for Eq. (\ref{16Juan}) to hold), we must have $B$ equal to
$c_{0}^{2}$.

At large values of $r$ the Newtonian limit is valid, so:

\begin{equation*}
g_{00}(r)\simeq c_{0}^{2} \left( 1-\dfrac{2 \phi}{c_{0}^{2}} \right) =c_{0}^{2}\left( 1-\dfrac{2GM}{c_{0}^{2}r}\right).
\end{equation*}

We must then have $A=GM$, so that we have for $g_{00}$, $g_{rr}$:

\begin{equation}
g_{00} =c_{0}^{2}\left( 1-\dfrac{2GM}{c_{0}^{2}r}\right); \quad g_{rr} =\dfrac{c_{0}^{2}}{g_{00}}
\label{18Juan}
\end{equation}
which are the well-known values of these coefficients for Schwarzschild metric.

We have said before that the quantity that can be called the ``gravitational potential'' in
GR is  $\dfrac{c_{0}^{2}}{2} \ln g_{00}$, in the sense that minus its gradient is equal to
the force on a static unit of mass. To study the difference of this potential with respect
to the Newtonian counterpart, it is covenient to express $\vec{g}$ in the following form:

\begin{equation}
\vec{g}=-\dfrac{\partial \ln g_{00}^{1/2}}{\partial \vec{x}}=-\dfrac{1}{g_{rr}^{1/2}}\dfrac{\partial}{\partial r}\ln g_{00}^{1/2}\vec{e}_{r}=c_{0}\dfrac{\partial g_{00}^{1/2}}{\partial r}\vec{e}_{r}
\label{19Juan}
\end{equation}

where $\vec{e}_{r}$ is the unit vector in the $r$ direction and where Eq. (\ref{16Juan}) has
been used. We see from Eq. (\ref{17Juan}) that if $g_{rr}$ was not affected by the
gravitational field, remaining equal to 1, we would obtain:

\begin{equation*}
g_{00}^{1/2}=c_{0}\left( 1 - \dfrac{GM}{c_{0}^{2}r}\right)
\end{equation*}

and $\vec{g}$ would be given by the Newtonian expression. It is then clear that what causes
$\vec{g}$ to differ from the Newtonian case is the fact that the gravitational field affect
the value of $g_{rr}$, causing it to be , in the present case, equal to
$g_{00}^{-1}c_{0}^{2}$.

For the vacuum case with non-zero cosmological constant, we need using Eq. (\ref{13Juan}) , that, noting that a $\lambda$ term can formally be treated like a homogeneous fluid with equation of state $p=- \rho$, takes the form:
\begin{equation*}
\nabla^{2}g_{00}^{1/2}=-g_{00}^{1/2}\lambda
\end{equation*}

Using again relationship (\ref{16Juan}), we find:
\begin{equation*}
\dfrac{\partial}{\partial r} \left(  r^{2} \dfrac{\partial g_{00}}{\partial r} \right)= -2 \lambda c_{0}^{2}r^{2}.
\end{equation*}

Integrating this equation with the same conditions as in $\lambda =0$ case, one obtains immediately:
\begin{equation*}
g_{00}=c_{0}^{2}\left(  1- \dfrac{2MG}{c_{0}^{2}r} - \dfrac{\lambda r^{2}}{3}\right), \quad g_{rr}=c_{0}^{2}g_{00}^{-1}
\end{equation*}

where  $M$ is a constant. This is the solution for the metric of a ``point'' mass in a
non-zero $\lambda$ vacuum, that was first obtained by Eddington in 1923\cite{Eddington}.

The Reissner-Nordstrom metric can be derived in an equally simple manner as we shall show in
a future work.

Assuming in the general spherical case that  relationship (\ref{16Juan}) holds, leads, together with Eq. (\ref{13Juan}), amongst other inconsistencies, to the fact that the mass of the distribution (for bounded distributions) depends on the enclosed mass and pressure-volume, that as we shall show in next section it is in contradiction with basics principles. It is therefore clear that this relationship it is not valid, a result that could have been anticipated by more direct (but more involved) argument.

To obtain the metric in the general case we shall use the results obtained in the vacuum
case together with the assumption that to obtain the field at $r$ only the matter
distribution up to $r$ is needed. This is a consequence of Birkhoff Theorem (REFERENCIA),
derived from G.R. This fact can be proved using Eq. (\ref{13Juan}) and some general
considerations, but we will simply take it here as a plausible assumption .This assumption,
that is simply the supposition that the full gravitational theory will also satisfy Newton's
iron sphere theorem, it is so natural that a derivation containing it can still be
considered to be based on basic principles. With the mentioned assumption and using Eq.
(\ref{18Juan}) it is clear that we must have for $g_{rr}$:

\begin{equation}
g_{rr}(r)=\left(1-\dfrac{2GM(r)}{c_{0}^{2}r}\right)^{-1}
\label{Juan20}
\end{equation}

where $M(r)$ is the mass enclosed within $r$.$M(r_{0})$ is defined so that if there were no
matter beyond $r_{0}$, the asymptotic Newtonian potential would be:

\begin{equation*}
\dfrac{GM(r_{0})}{r}.
\end{equation*}

In next section we shall show from basics considerations that $M(r)$ is given by:

\begin{equation}
M(r)=\dfrac{4 \pi}{c_{0}^{2}}\int^{r}_{0} \rho(r^{\prime})r^{\prime 2}dr^{\prime}
\label{Juan21}
\end{equation}
where $\rho$ is the energy density.Using Eq. (\ref{Juan21}) in Eq. (\ref{Juan20}) the known result for $g_{rr}$ is recovered\cite{Landau2}. Inserting this expression for $g_{rr}$ in Eq. (\ref{13Juan}) we obtain an equation for $g_{00}$
\begin{equation}
\dfrac{1}{g_{rr}^{1/2}r^{2}}\dfrac{\partial}{\partial r} \left( \dfrac{r^{2}}{g_{rr}^{1/2}} \dfrac{\partial g_{00}^{1/2}}{\partial r} \right)=\dfrac{4 \pi G}{c_{0}^{4}}\left( \rho + 3P \right)g_{00}^{1/2}
\label{Juan22}
\end{equation}
with the conditions:

\begin{equation*}
\dfrac{\partial g_{00}^{1/2}}{\partial r}\mid_{r=0}=0; \quad \quad g_{00}(\infty)=c_{0}^{2}
\end{equation*}

However, rather than integrating this equation, we shall obtain the solution at the end of
next section by a more physicaly meaningful procedure. 

Once we have the metric, the trajectories of free falling particles can be obtained using
the geodesic equation in space-time for this metric, which is the standard procedure in
textbook on GR. However, to end this section, we shall indicate how this can be done in a
simple manner and in much closer keeping with classical mechanics. 

GR implies very small corrections for planetary orbits and we think it is instructive to
describe the effects by mean of small correcting terms added to the Newtonian equations and
explain their origin. To say that in GR gravity is not a force, that  planets merely follow
space-time geodesics and, in consequence, use a qualitatively different treatment to deal
with quantitatively very small effects, does not seem to us the most convenient
presentation.

In fact, the very description of gravity as the space-time geometry is an adventitious
interpretation and not at the essence of Einstein's theory, as pointed out by
Weinberg\cite{Weinberg}. This interpretation may be interesting and compelling for those who
already know the theory, but to use it to convey to the layman the meaning of GR does not
seems the most expedient way.

The Lagragian for a particle in a static gravitational field may be written in the form:

\begin{equation}
L=-m \dfrac{ds}{dt}=-mc_{o}^{2}\sqrt{g_{00}-\left(\dfrac{dl}{dt}\right)^{2}};
\label{Juan23}
\end{equation}
\begin{equation*}
d^{2}l\equiv g_{\mu \nu}dx^{\mu}dx^{\mu}
\end{equation*}

where $s$ is the invariant relativistic interval and $l$ is the arc length ($\mu , \nu$ run
from 1 to 3).

If space was Euclidean, one can easily show that the equation of motion derived from Eq.
(\ref{Juan23}) would be:

\begin{eqnarray}
&\dfrac{d\vec{v}}{dt}&=  \left( -\dfrac{1}{2}\vec{\nabla}g_{00} \right)_{\perp}  +\left( -\dfrac{1}{2}\vec{\nabla}g_{00} \right)_{\parallel} \left( 1-2\left( \dfrac{v}{c}\right)^{2} \right)=-\dfrac{1}{2}\vec{\nabla}g_{00}+\dfrac{1}{c^{2}}\left( \vec{v} \cdot \vec{\nabla}g_{00}  \right) \vec{v}; \\
 &\vec{v}& \equiv \dfrac{d \vec{x}}{dt} \nonumber
\label{Juan24}
\end{eqnarray}

where the suffixes $\perp$, $\parallel$  denotes respectively the components perpendicular
and parallel to $\vec{v}$. In this expression both $\vec{v}$ and its derivative are with
respect to coordinate time $t$. Using the local  proper time,  $\tau$ (not to be confused
with particle proper time), we have:

\begin{equation*}
\dfrac{d\vec{v}}{d\tau}= \vec{g}_{\perp} +\left(1-\left( \dfrac{v}{c_{0}}\right)^{2}\right)\vec{g}_{\parallel}
 \end{equation*}

\begin{equation}
\vec{g}= -\dfrac{c_{0}^{2}}{2g_{00}}\vec{\nabla}g_{00};\qquad \vec{v}\equiv\dfrac{d\vec{x}}{d\tau}
\label{Juan25}
 \end{equation}
 
where $\vec{g}$ is the ``gravitational field'' (i.e. the force on  a unit mass at rest).
This equations can be obtained immediately using the fact that in a locally inertial system
(i.e. free falling) the acceleration vanishes . Therefore, both Eq. (22) and Eq.
(\ref{Juan25}) merely express the simple (non-Riemannian) EP, which implies a velocity
dependent gravitational force. Note that from Eq. (23) it is obvious that $v$ will always
remain smaller than $c_{0}$, as it must be when time $\tau$ is used. However Eq. (22) do not
imply  such restriction, because in time $t$ $v$ can be larger than $c_{0}$, although it
have to be smaller than $c(\vec{x})$. 

In general, space it is not Euclidean, and an extra term must be included in Eq.
(\ref{Juan24}). In the spherically symmetric case we have: 

\begin{equation}
\left(\dfrac{dl}{dt}\right)^{2}=\left(\dfrac{d\chi}{dt}\right)^{2}+\left(r\dfrac{d\theta}{dt}\right)^{2},
\label{Juan26}
\end{equation}
\begin{equation*}
d\chi \equiv g_{rr}^{1/2}dr
\end{equation*}

where, since it is obvious that orbits must remain in a plane, we are considering only two
coordinates $\chi$ (or $r$) and $\theta$, that determine the position within that plane.

Using Eq. (\ref{Juan26}) in Eq. (\ref{Juan23}) one obtains the corresponding Lagrangian from
which the equations of evolution for $\chi$, $\theta$ can be obtained in the standard
manner. But  it is simpler and more enlightening to recall the classical derivation and
remark the differences. Classically we may write: 

\begin{equation}
\dfrac{d\vec{v}}{dt}=\dfrac{d}{dt}\left( \dot{r}\vec{e_{r}}+r\dot{\theta}\vec{e_{\theta}}\right)=\vec{F}
\label{Juan27}
\end{equation}

where $\vec{e_{r}}$, $\vec{e_{\theta}}$, are the unit vector along the radial and tangential
directions, and $\vec{F}$ is the total force per unit mass, that in the Newtonian case is
given simply by $\vec{g}$. Using the following relationship:

 \begin{equation}
 \dot{\vec{e_{r}}}=\dot{\theta}\vec{e_{\theta}}; \quad \dot{\vec{e_{\theta}}}=-\dot{\theta}\vec{e_{r}},
\label{Juan28}
\end{equation}
where the dot denotes derivation with respect to $t$, we have:
\begin{equation*}
\left( \ddot{r} - r \dot{\theta}^{2}\right)\vec{e_{r}}+\dfrac{ \dfrac{d}{dt}\left( r^{2} \dot{\theta}\right)}{r}\vec{e_{\theta}}=\vec{g}.
\end{equation*} 

To obtain the evolution equations in the present case we only need to note that the
underlying reasons for Eq. (\ref{Juan28}) are the relationships satisfied in the Euclidean
plane. The relationships satisfied in a non-Euclidean plane with radial symmetry around
$r=0$ can be obtained by equally simple geometrical considerations:

\begin{equation*}
 \dot{\vec{e_{r}}}=\left(\sin \rho \right)\dot{\theta}\vec{e_{\theta}}; \quad \dot{\vec{e_{\theta}}}=- \left( \sin \rho\right) \dot{\theta}\vec{e_{r}}.
\end{equation*}

\begin{equation}
\sin \rho \equiv \dfrac{dr}{d\chi}=g_{rr}^{-1/2}.
\label{Juan29}
\end{equation}

 Representing the actual non-Euclidean plane as an axialy symmetric curved plane embedded in
Euclidean space and being tangent to the plane $z=0$ at the origin, $\rho$ is the angle
between the radial direction within the curved plane and the $z$ axis.

In the present case it also holds that the derivative of $\vec{v}$ with respect to
coordinate time $t$ is equal to the total force per unit mass, $\vec{F}$, but now $\vec{v}$
is given by:

\begin{equation*}
\vec{v}= \dot{\chi}\vec{e_{r}}+ r \dot{\theta}\vec{e_{\theta}},
\end{equation*}

and $\vec{F}$ is given by the left hand side of Eq. (22). Using 	 Eq. (\ref{Juan29})
for the derivative of the base vector, we finally find

\begin{equation}
\left( \ddot{\chi} - g_{rr}^{-1/2} r \dot{\theta}^{2}\right)\vec{e_{r}}+\left[\dot{r}\dot{\theta}+ \dfrac{d}{dt}\left( r \dot{\theta} \right) \right]\vec{e_{\theta}}=-\dfrac{1}{2}\vec{\nabla}g_{00}+\dfrac{1}{g_{00}}\left(\vec{\nabla}g_{00} \cdot \vec{v}\right)\vec{v},
\label{Juan30}
\end{equation}
\begin{equation*}
\vec{v}\equiv \dot{\chi}\vec{e_{r}}+r\dot{\theta}\vec{e_{\theta}}.
\end{equation*}

From the radial and the tangential part of these equation we obtain respectively:
\begin{equation*}
g_{rr}^{-1/2} \dfrac{d}{dt} \left( g_{rr}^{3/2}\dot{r}\right)-r^{2}\dot{\theta^{2}}=-\dfrac{1}{2}\dfrac{\partial g_{00}}{\partial r}; \quad \quad \dfrac{d}{dt} \left( \dfrac{r^{2}\dot{\theta}}{g_{00}}\right)=0
\end{equation*}

where Eq. (\ref{16Juan}) has been used, so, at variance with Eq. (28), that is valid for any
spherically symmetric metric, the first of these equations is only valid for vacuum metrics.
These equation can be integrated exactly, but we will only discuss simple cases that can
easily be treated approximately in a more transparent manner.

For almost circular orbits, neglecting terms proportional to $\dot{r}$, we have:

\begin{equation}
g_{rr}^{-1/2}\ddot{r}-g_{rr}^{-1/2} r\dot{\theta}^{2}=-\dfrac{1}{2}\vec{\nabla}g_{00}
\label{Juan31}
\end{equation}

\begin{equation*}
\dfrac{d\left(\dfrac{r^{2}\dot{\theta}}{g_{00}}\right)}{dt}=0  
\end{equation*}

The differences with respect to the  classical case are three fold: First there is the fact
that the force per unit mass at rest, $\vec{g}$,  is not exactly equal to the Newtonian one.
Then we have the velocity dependence of the gravitational force and, finally we have the
non-Euclidean character of space, which is fully described by the function $r(\chi)$.
However, even when $\vec{g}$ (related to the acceleration in time $\tau$) differs from the 
Newtonian one, what appears in the right hand side of Eq. (\ref{Juan30}) (which correspond
to time $t$) is exactly equal to the Newtonian case, on the other hand, as we have said, for
almost circular orbits the radial velocity dependence is negligible,but not the tangential
one, that renders the precensece of $g_{00}$ in the last of Eq. (29).So, the difference with
respect to the Newtonian case are due to the velocity dependence of $\vec{F}$ and to the
non-Euclideanity of space.

 The precesion of the perihelion of almost circular orbit can easely be obtained.
Integrating the last of Eq. (29) and inserting it in the first, we find:

\begin{equation*}
\ddot{r}=g_{rr}^{-1} \left( r \dot{\theta}^{2}- \dfrac{1}{2}\dfrac{\partial}{\partial r}g_{00}\right)=g_{rr}^{-1}\left( \dfrac{J^{2}g_{00}^{2}}{r^{3}}-\dfrac{GM}{r^{2}}\right)
\end{equation*}

where $J$ is an integration constant. For a small displacement, $\Delta r$, from  the value
of $r$ at  which the last parentheses vanishes, $r_{0}$, we have:

\begin{equation}
\ddot{r}\simeq -g_{rr}^{-1} \left(\dfrac{GM}{r_{0}^{3}}\left( 1-\dfrac{4GM}{r_{0}c_{0}^{2}}\right) \right)\Delta r = -\omega_{r}^{2}\Delta r
\label{primera}
\end{equation}

where $\omega_{r}$ is the angular frequency of small radial oscillations. For the angular
motion we have:

\begin{equation*}
\omega_{\theta} \equiv \dot{\theta}= \left( \dfrac{GM}{r^{3}_{0}}\right)^{1/2}.
\end{equation*}

Therefore, the angular frequency of the perihelion, $\Omega$, is given by:
\begin{equation}
\Omega= \omega_{r}-\omega_{\theta} \simeq \dfrac{3GM}{r_{0}c_{0}^{2}}\omega_{\theta}.
\label{segunda}
\end{equation}

It must be noticed that when the orbit has finite eccentricity, $r_{0}$ is not the mean
radius. From its definition:

\begin{equation*}
r_{0}=\dfrac{J^{2}}{GM}=a\left(1-e^{2} \right)
\end{equation*}

where $a$ is the semi-major axis and $e$ is the eccentricity. For finite radial oscilations
higher order terms on $\Delta r$ must be included in Eq. (31), but we know that without the
relativistic corrections $\omega_{r}$ and $\omega_{\theta}$ (that for general orbits is
given by the above expression with $r_{0}=a$) remain equall. Therefore, with this value of
$r_{0}$, the above expression for $\Omega$ is exact to first order in the potential. Notice
that a negative value of $\Omega$ means that the perihelion moves in the direction of the
revolution. From the above computation it is clear that $2/3$ of $\Omega$ are due to the
velocity dependence of $\vec{F}$, and this is the value of $\Omega$ corresponding to
Einsteins's 1912 theory. The remaining third comes from the presence of $g_{rr}$ in Eq.
(\ref{primera}), obviously related to the non-Euclidianity of space. It is interesting to
note that in Nordstrom theory, where space-time is comformally flat (therefore, $g_{rr}$ is
the inverse of that for GR), the spatial effect has the opposite sign to that for GR, while
the effect of the velocity dependence of $\vec{F}$ is the same. However, in that theory, the
gravitational potential do not enter Eq. (28) as in the Newtonian theory. The sum of all
effects render a value o $\Omega$ that is in magnitude $7/3$ of that for GR, bu with the
opposite sign.

The deflection of light rays can also be easily computed. From the derivation of Eq.
(\ref{Juan30}) it is clear that in quasi-Cartesian coordinates and to first order  on $G$ we
have:

\begin{equation}
\dfrac{d \vec{V}}{dt}= -\dfrac{1}{2}\vec{\nabla}g_{00}+ \dfrac{1}{2}\left(\vec{\nabla}g_{00}\right)_{\parallel} - \left[  \dot{\theta} \dot{\chi}\left( g_{00}^{-1/2}-1 \right)\vec{e_{\theta}}-r \dot{\theta^{2}} \left(  g_{00}^{-1/2}-1\right)\vec {e_{r}} \right]
\label{Juan?}
\end{equation}

where we have used the fact that for light $v=c$ and where the last parentheses comes from
the fact that  replacing Eq. (\ref{Juan28}) by  (\ref{Juan29}) is equivalent  to adding the
apparent force given by the parentheses while retaining the Euclidean affinity structure. We
have expressed the parentheses in polar coordinates because it is simpler to derive, but now
must be transformed to Cartesians. Considering a ray with impact parameter $r_{0}$ and
taking the coordinate system so that the ray moves in the $x$, $y$ plane along the straight
line (almost) $y=r_{0}$, we then have:

\begin{equation*}
\dfrac{d V_{y}}{dt}=-\dfrac{GM}{r^2}\dfrac{r_{0}}{r} -\dfrac{GM}{r^2}\dfrac{r_{0}}{r}
\end{equation*}

where the first term on the right hand side corresponds to the homologous term in Eq.
(\ref{Juan?}) and the last one corresponds to the parenthesis. The second term in Eq.
(\ref{Juan?}) does not contribute because it goes in the x direction. Taking $t=0$ at the
moment of maximum approach and assuming that light traces a straight line at constant speed,
$c_{0}$ (zeroth order on $G$), it is obvious that the deflection,  $\Delta \phi$, is
given by:

\begin{equation*}
\Delta \phi= \dfrac{2}{c_{0}} \int^{\infty}_{must be0} \dfrac{dV_{y}}{d t}dt=\dfrac{2}{c_{0}}\int^{\infty}_{0} \dfrac{2GM}{(r_{0}^{2}+ c_{0}^{2}t^{2})^{3/2}}dt=\dfrac{4GM}{r_{0}c_{0}^{2}}
\end{equation*}

Which is the well-known result, that is usually  derived in a rather different manner. We
have seen that one half of the effect comes from the non-Riemannian EP (first term of the
right in Eq. (\ref{Juan?})) while the other half comes from the non-Euclideanity of space,
(last parentheses in Eq. (\ref{Juan?}).

\section{Historic overtones and other topics}

We have said before that the argument that led Einstein to reject Eq. (\ref{6Juan}) was
incorrect. The argument\cite{A.Enstein1} is based in the demonstration of the 
non-conservation the momentum of a distribution of particle interacting through a
gravitational field satisfying Eq. (\ref{6Juan}) and following the dynamic that he had
previously derived\cite{A.Enstein2}. Met with this contradiction Einstein chose to reject
Eq. (\ref{6Juan}), but the problem lied not with this equation, that we have seen that is
correct (for negligible pressure), but with the assumption that space can remain flat in a
gravitational field. We have mentioned this fact before; here we shall explicitly show it in
a simple manner. To this end, instead of the general case, we shall consider the case of two
interacting point masses. Furthermore, we assume that both masses are instantaneously at
rest. To obtain the acceleration of the masses we may use Eq. (\ref{Juan24}). This
expression does not fully incorporate non-Euclidianity, since Eq. (\ref{Juan28}) rather than
Eq. (\ref{Juan29}) are  implicit, but this is irrelevant for motion in a radial direction.

\begin{equation}
\dfrac{d\vec{v}}{dt}=-\dfrac{1}{2}\vec{\nabla}g_{00} \quad \quad for \quad \vec{v}=0
\label{32Juan}
\end{equation}

Now from Lagrangian Eq. (\ref{Juan23}) it is clear that the linear momentum is given by:

\begin{equation*}
\vec{p}=\frac{mc_{0}\vec{v}}{c \sqrt{1-( \dfrac{v}{c})^{2}}}.
\end{equation*}

This agrees with the expression derived by Einstein\cite{A.Enstein2}. It is then clear that
the change of momentum of mass 1 and 2 are respectively:

\begin{equation*}
\dfrac{d\vec{p}_{1}}{dt}=\dfrac{m_{1}c_{0}}{c_{1}}\dfrac{d\vec{v}_{1}}{dt}=-\dfrac{m_{1}c_{0}}{2c_{1}}\vec{\nabla}c_{1}^{2}; 	\quad \dfrac{\vec{dp}_{2}}{dt}=-\dfrac{m_{2}c_{0}}{2c_{2}}\vec{\nabla}c_{2}^{2}
\end{equation*}

where $c_{1}$, $c_{2}$ are the values of the $c$ field ($\equiv g_{00}^{1/2}$) at mass 1 and
mass 2 respectively.  For the gradient we have:

\begin{equation*}
\vec{\nabla}c^{2}_{1}=\dfrac{1}{g^{1/2}_{rr_{1}}}\dfrac{dc_{1}^{2}}{dr}\vec{e}_{r}; \quad \vec{\nabla}c_{2}^{2}=-\dfrac{1}{g_{rr_{2}}^{1/2}}\dfrac{dc^{2}_{2}}{dr} \vec{e}_{r},
\end{equation*}
\begin{equation*}
r \equiv \vert \vec{r}_{2}-\vec{r}_{1} \vert
\end{equation*}

where $\vec{e}_{r}$ is a unitary vector in the direction of $\vec{r}$ ($\equiv 
\vec{r}_{2}-\vec{r}_{1}$). Using Eq. (\ref{18Juan}) for $c$ ($\equiv g_{00}^{1/2}$) and
$g_{rr}$ ($= c^{2}_{0}/g_{00}$), we find:

\begin{equation*}
\dfrac{d\vec{p}_{1}}{dt}=m_{1}\dfrac{dc_{1}^{2}}{dr} \vec{e}_{r}=\dfrac{m_{1}m_{2}G}{r^{2}}\vec{e}_{r}; 	\quad \dfrac{d\vec{p}_{2}}{dt}=-\dfrac{m_{1}m_{2}G}{r^{2}}\vec{e}_{r}.
\end{equation*}

The change of the total momentum, therefore vanishes. The analysis of this simple example
makes it clear what was wrong with Einstein`s 1912 theory. In the above expression $\vec{r}$
is the actual ``metric'' velocity (i.e. $v=\dfrac{dl}{dt}$, $l$ being the length of arc). In
polar coordinates in the plane of motion:

\begin{equation*}
v=\sqrt{g_{rr}\dot{r}^{2}+ \left(r \dot{\theta} \right)^{2}}
\end{equation*}

In  Einstein's theory space was flat, so in the present case ($\dot{\theta}=0$) $v$ is
identical to $\dot{r}$. But in this case instead of Eq. (\ref{32Juan}) we would have

\begin{equation*}
g_{rr}^{1/2} \dfrac{d \vec{v}_{E}}{dt}=-\dfrac{1}{2}\vec{\nabla}g_{00}
\end{equation*}

where $\vec{v_{E}}$ is $\vec{v}$ in Einstein theory (i.e. $\vert \vec{v}\vert =
\dot{r}$).Then we would have:

\begin{equation*}
\dfrac{d\vec{p}_{1}}{dt} \mid_{E}=\dfrac{1}{g_{rr_{1}}^{1/2}}\dfrac{c_{0}m_{1}m_{2}G}{r^{2}}\vec{e}_{r}
\end{equation*}

and a similar expression for mass 2. This is clearly incompatible with momentum
conservation.

As we have seen, confronted with this contradiction Einstein chose to retain the
Euclidianity of space and to modify Eq. (\ref{6Juan}) so as to make it compatible with
momentum conservation (and Euclidianity) although it was then incompatible with the strict  
(i.e. for any field strength) EP.

When sometime latter he realized that space could not remain Euclidean in a gravitational
field, he retook the EP, that  he had rejected very reluctantly, and started a new path
through differential geometry. No discussion and revision of the previous work at the new
light can be found in the literature. This may seem quite reasonable: since the
non-Euclidianity of space in a gravitational field turns the spatial metric coefficients,
$g_{\mu \nu}$, into some sort of gravitational potentials, obtaining an equation for just
$g_{00}$ does not look very interesting. But before going after the full theory Einstein
could as well have clarified the situation with respect to Eq. (\ref{6Juan})  and see how
far he could  have gone with it in some simple cases. In fact, as we have shown here, if he
had followed that course and realized of the argument relating $g_{rr}$ and $g_{00}$ in the
vacuum case, Schwarzschild's metric and its consequences would have been around since 1912.
The mentioned argument is akind to  the one given by Einstein for showing the
non-Euclidianity of space in a rotating system, only that somewhat more sophisticated
because it involves comparing spatial geometries in flat space-time and curved space-time.
Thus, we think that this could have been a realistic course of events. Had this been the
case, not only would have been available several of the most relevant results of GR for
astrophysics since that early date, but substantial insight would have been added to the
full theory when it became available,  by showing its direct connection with basic principle
through the analysis of  simple cases. Furthermore, we think that, after realizing that
space could not remain flat in a gravitational field, Einstein initial project of dealing
with the static and stationary case could have been fulfilled. It is true that the task then
would has exceeded that of the original Einstein intentions, since equations  must be obtain
to determine all metric coefficients, not just $g_{00}$. However, that task would have been
still much smaller than obtaining the full G.R. equations and, again, would have contributed
substantial insight into it. We have not so far pursued that course, but we now briefly
describe here how we think it could be done.

 We have said that Einstein found that Eq. (\ref{6Juan}) can not be derived from a
variational principle (so that energy-momentum conservation is not guaranteed), but we know
that that equation is correct (for negligible pressure) and, of course, momentum is
conserved. How are this two fact made compatible by a curved space?. Once it is recognized
that the gravitational field affect the space metric, the corresponding metric coefficients
become dynamical quantities and, therefore, must be included in the  Lagrangian for matter
and the gravitational field. The additional term in the Lagrangian must be invariant under
spatial coordinate transformation, so it must  be formed out of $c$ (that is constant under
spatial transformations) and the Ricci scalar of the spatial metric, $R$. Simple
considerations lead to a term of the form $ARc$, with $A$ a constant. This constant must be
chosen so that Eq. (\ref{6Juan}) is obtained throug variation of the Lagrangian with respect
to $c$. Variation with respect to spatial metric coefficients  should provide the other
equations needed. On the other hand, from the consideration of the flat space-time metric in
a rotating system\cite{Landau3} it may readily be shown that the term :

\begin{equation*}
-2\dfrac{\Omega^{2}}{c^{3}}
\end{equation*}

where $\Omega$ stand for the rotation speed, must be added to the right hand side of Eq.
(\ref{6Juan}) in the presence of rotation (i.e., in a non globally synchronizable system).
Now, the rotation velocity $\vec{\Omega}$ is related to the coefficients $g_{0 \mu}$ ($\mu$
from 1 to 3) by\cite{Landau4}

\begin{equation}
\vec{\Omega}= \dfrac{c_{0}}{2} curl \vec{g}; \quad \quad \vec{g}\vert_{\mu}\equiv - \dfrac{g_{0\mu}}{c^{2}_{0}}.
\label{33Juan}
\end{equation}

Inferring the term that must be added to the Lagrangian to obtain the modified Eq.
(\ref{6Juan}) and variating the Lagrangian  with respect to all metric coefficients, the ten
needed equations could be obtained.

This line of development might have difficulties, in particular, it must be noted that the
spatial metric coefficients enter the Lagrangian through derivatives of up to the second
order ( this is true in GR for all metric coefficients). However, it seem, in principle, a
reasonable path to follow towards GR and a clarifying analysis once the theory is available.
If this approach worked, space-time geometrical formalism would only be needed to deal with
time dependent gravitational fields.

We have seen that Einstein obtained Eq. (\ref{6Juan}) from simple considerations and that GR
leads to Eq. (\ref{13Juan}) which agrees with Eq. (\ref{6Juan}) only when pressure  is
negligible. The question now is what was missing in the arguments leading to Eq.
(\ref{6Juan}) and whether Eq. (\ref{13Juan}) could be derived through some simple
consideration that provides that missing element. This question is related to the
interesting and very delicate issue of the gravitational effect of pressure and given its
relevance for this work we consider it convenient to pay some attention to it.

The source of the gravitational field in GR is the energy-momentum tensor, $T_{ij}$ (with
$i,j$ from 0 to 3), whose purely spatial components corresponds to pressure (the stress
tensor).It is  then no surprise that pressure affect the metric tensor, $g_{ij}$, which are
in some sense a set of ten gravitational potentials (they are all needed to determine
particle trajectories). However, as we stated before,  the coefficient $g_{00}$ is the one
that provides the force on a unit mass at rest, so it is in close relationship with the
Newtonian potential and the fact that pressure contribute to its sources, as seen in Eq.
(\ref{13Juan}), seems rather buffling. In one of his 1912 papers\cite{A.Enstein1}, Einstein
discussed the possibility  of the gravitational action of stresses (i.e. its ability to
weight in a given gravitational field) and rejected it with an argument that involved the EP
and the equivalence of mass and energy. He seems to have assumed that what he has proved
correctly for the passive coupling of matter to gravity was also true for the active
coupling (i.e. the ability of matter to generate a gravitational field). In the Newtonian
theory the superposition principle holds and this implies that momentum conservation can
only be satisfied if it holds for any couple of interacting elements (i.e. mass points or
volume elements, for extended systems), that implies the action and reaction law. In this
case it is clear that the active and passive coupling must be equal. In the present case,
however, the superposition principle does not hold and, therefore, the equality of action
and reactions is not satisfied. This explains that we may have a passive coupling
proportional to $\rho$ while the active coupling is proportional to $\rho +3P$. But for
stable and bounded objects interacting at large distances, so that the superposition
principle holds, Einstein's argument applies also to the active coupling, implying a net
gravitational effect proportional to the total energy of those objects.

 Einstein considered a box filled with radiation, which is an stable bounded system and,
therefore, as he had shown, its gravitational effect at distances much larger than its size
depends solely on the sum of the masses in the box (for a small box i.e. negligible binding
energy). But, having concluded correctly that the net gravitational effect of stresses has
to vanish for this system, he arrived at the wrong conclusion (al least implicity) that the
gravitational effect of stresses must vanish (which is exactly true only for its passive
role), because he only considered  the stresses in the containing walls and not the negative
stresses within it, associated with the radiation pressure. In fact, Einstein argument, that
we shall see later in more detail, is compatible with stresses contributing to $g_{00}$,
because in any stable bound system positive an negative stresses cancel each other. In self
gravitating systems with positive pressure the stabilizing stresses are provided by the
gravitational field (the energy-momentum tensor of gravity is another delicates issue, but
we can not address it here).

Knowing  that there is not a clear argument to exclude pressure from  contributing to the
gravitational field strength, the question is whether a simple and direct argument can be
given to derive Eq. (\ref{13Juan}). The answer to this question could be that, after all,
the simplest version of GR (with only linear term on the Ricci scalar on the  Lagrangian and
without cosmological term) follows entirely from the strong EP and Eq. (\ref{13Juan})
followes from it. Thus, the train of thought leading to the presence of the term $3P$ in the
Eq. (\ref{13Juan}) could be considered to be the argument asked for. Because of its
interest, we shall review that train of thought. However this scarcely can be considered a
direct argument like  that leading to Eq. (\ref{6Juan}), because it goes all the way through
GR. We know of a simple argument showing the gravitational effect of pressure, but it
involves a non-static field and we prefer to present it in a future work. Einstein, on the
other hand, had no reason in 1912 to search for that argument, and, to our knowledge, it was
only on completing GR that he realized of the gravitational effect of pressure in the sense
discussed here.

To understand the origin of the $3P$ term in Eq. (\ref{13Juan}) it is interesting to note
that  in a preliminary theory, where the field equations were:

\begin{equation}
R_{ij}=\dfrac{8 \pi G}{c_{0}^{2}}T_{ij}
\label{34Juan}
\end{equation}

the equation for $g_{00}$ in the static field would be given by Eq. (\ref{6Juan}) rather
than by Eq. (\ref{13Juan}). Therefore, it is clear that the key to the questions that we are
analysing lies with the origin of the extra term in the right hand side of Eq.
(\ref{7Juan}). This equation is equivalent to:

\begin{equation}
R_{ij}^{*}\equiv R_{ij}-\dfrac{1}{2}g_{ij}R=\dfrac{8 \pi G}{c_{0}^{2}}T_{ij}.
\label{35Juan}
\end{equation}

Our question has now been reduced to explaining why it is $R_{ij}^{*}$ rather than $R_{ij}$
that must appear in the left hand side of the field equations. This equations follows
outright by variating the simplest GR Lagrangian, but in order to see the connection of our
question with physical principles we shall follow the usual explanation, although with a
somewhat different presentation.

 Both the metric coefficients and the components of the energy-momentum tensor are sets of
ten algebraically independent quantities. However, since the choice of coordinates is
arbitrary and fixing a coordinate system involves four independent functions (the
coordinates conditions), only six of the $g_{ij}$ and $T_{ij}$ are functionally independent.
That is, once the coordinates have been fixed, all possible energy-momentum distributions
can be described by six independent functions. Any sets of ten functions can describe an a
priori possible $T_{ij}$ distribution, but it would correspond to certain intrinsic
distribution as represented in certain coordinate system. When the latter has been fixed,
not all set of ten functions are possible $T_{ij}$ distribution. In other words, the
intrinsic structural possibilities of both $g_{ij}$ and $T_{ij}$ are span by sets of six
independent functions. The same applies to $R_{ij}$, which can be expressed in term of the
$g_{ij}$. In consequence, we should have just six quantities intrinsically characterizing
$R_{ij}$ and $T_{ij}$, but since on any given system of coordinates there are ten $R_{ij}$
and $T_{ij}$, the field equations must be ten equations, although only six must be
functionally independent. However, since the $g_{ij}$ and $T_{ij}$ are different structures
(in fact $T_{ij}$ is not even specified in terms of the constituting field), the only way in
which the number of independent equation can be reduced to six is by having the side of the
equation corresponding to gravity, which is geometrical in character and explicitly
specified in terms of its constituting fields ($g_{ij}$), satisfying four identities. But
from Bianchi's identities it is known that:

\begin{equation*}
R_{\quad j;i}^{*i}=0.
\end{equation*}

$R^{*}_{ij}$ must then be the ``geometrical'' side of the field equations, as shown in Eq.
(\ref{35Juan}). Taking the four divergence on both sides of Eq. (\ref{35Juan}) we
immediately have:

\begin{equation}
T_{\quad j;i}^{i}=0
\label{36Juan}
\end{equation}

that expresses energy-momentum conservation, that appears now as a direct consequence of
general covariance (i.e. the laws of physics has the same form in all coordinated
systems).We remind, however, that general covariance has no physical content, which is
provided by the fact that general covariance,  a conventional requirement that should be met
by any theory, can be achieved with just matter and gravity, without foreign elements, a
fact  that follows from the strong EP. If this principle did not hold and certain additional
fields caused the laws of physics (matter and gravity) not to be the same in all free
falling systems, the argument given above would also be valid, but instead of Eq.
(\ref{36Juan}) we would have the divergence of $T_{ij}$ plus a tensor build up  from the
foreign fields set equal to zero. In this case, the conservation that followed from general
covariance (that  holds by construction) would not imply the local  conservation of
energy-momentum (ordinary) alone.

In some works one can see the above argument shortened  to saying that, given the empirical
fact described by Eq. (\ref{36Juan}), the left hand side of the field equations must have
identically vanishing divergence, so it must be $R^{*}_{ij}$. This is a complete reversal 
of the logic of the argument given here, in which the fact of having $R^{*}_{ij}$ on the
left hand side of Eq. (\ref{35Juan}) followed from general covariance, while energy-momentum
conservation follows from Eq. (\ref{35Juan}), and it is not a valid reasoning. If energy
-momentum conservation were merely an empirical  fact or consequence of another symmetry
(i.e. other than general covariance), Eq. (\ref{34Juan}) could be valid. We should then 
have:

\begin{equation*}
R^{i}_{\quad j;i}=0
\end{equation*}

which is not an identity, but there would be no logical impediment to it. Eq. (35),
althouggh it satisfies general covariance, it does not follow from a variational principle
and, therefore, no conservation law is implied in general by that property. There is a local
conservation of matter energy-momentuml, that was already there, but not an ordinary
conservation of ``total'' energy-momentum (including gravity). In fact, it was demanding
that this be the case (which follows automaticaly from a variatiional principle) that
Einstein obtained Eq. (34)

We have seen that the $3P$ term in Eq. (\ref{13Juan}) is a  consequence of general
covariance, or rather, of the fact, implied by the strong EP, that it can be achieved with
only matter and gravity. Now we shall use Einstein argument showing that only mass
gravitates (valid for stable bounded objects) to obtain an expression for the mass of an
static and bounded spherical distribution of energy-momentum. 

Considerer an static and bounded distribution of mass and pressure. It is clear that at
large enough distances the gravitational potential takes the form:

\begin{equation}
\phi \sim \dfrac{GM}{r}.
\label{37Juan}
\end{equation}

The constant $M$ we call the mass of the distribution. By considering the acceleration
experienced by the whole system within a uniform gravitational field and using the EP
Einstein showed that  the gravitational mass of the system,$M$, must be equal to the
inertial mass, that is, it is the latter that couples passively to the gravitational field
($g_{00}$) and the same is true for the active coupling when computing weak fields. Note
that this argument does not apply to a portion of the system. A volume element, $dV$,
couples actively to $g_{00}$ not through its mass, but through its mass plus $3P$ times
$dV$, as we discussed earlier. Now, as pointed out by Einstein, it is clear that the total
energy and momentum of the system must depend on the velocity of its center of mass like a
mass point in special relativity. This is true even when the latter theory does no hold in a
region around the system where the field is sufficiently strong and it  can be proved
rigorously by considering energy-momentum conservation for a set of systems that interact 
weakly (the field on any system due to all other systems being weak) without changing their
inner structure. From this fact, the proportionality between mass (inertial) and total
energy follows immediately. In consequence, $M$ in Eq. (\ref{37Juan}) must be given by the
total energy of the system. It would be tempting to write:

\begin{equation}
M=\int \rho dV +\dfrac{1}{2 c_{0}^{2}}\int \left(\dfrac{g_{00}^{1/2}}{c_{0}}-1 \right) \rho dV
\label{38Juan}
\end{equation}

where the second term is the gravitational binding energy.However, writing the binding
energy as one half of the total gravitational energy (sum of the gravitational energy of all
mass elements) is only an approximation, since the presence of $c$ on the right hand side of
Eq. (\ref{13Juan})(or Eq. (\ref{6Juan})) makes it clear that the superposition principle does
not hold (in consequence, the energy of mass $i$ due to mass $j$ is not symmetric in $i$,
$j$). But we shall see now that in the case of a spherically symmetric cloud $M(r)$ can be
obtained exactly using the vacuum case solution. This expression for $M(r)$ can be used to
obtain the full solution ($g_{00}$ and $g_{rr}$) in the non.homogeneous spherical case, as
we have shown in the previous section. If matter and pressure distributions beyond radii $r$
do not affect the solution within $r$, the value of $g_{rr}$ at $r$ must be given by Eq.
(18)

\begin{equation*}
g_{rr}(r)= \left( 1- \dfrac{2GM(r)}{c_{0}^{2}r}\right)^{-1}
\end{equation*}

with $M(r)$ the mass enclosed within $r$. The value of $g_{00}(r)$ when normalized so that
$g_{00}(0)=c_{0}^{2}$ is also unafected by matter and pressure outside $r$, but when, as
usual, is normalized so that $g_{00}(\infty)=c_{0}^{2}$, there is a dependence on outside
matter.To obtain $M(r)$, we shall assume that there is no matter beyond $r$ (which would be
irrelevant under our assumption) and compute the total work, $W$, that must be exerted on
the  cloud to disperse it  by taking layer after layer  to infinity (obviously, $W$ is
numerically equal to the binding energy). $W$ is clearly given by:

\begin{equation}
W(r)= - \int^{r}_{0} \left(\dfrac{g_{00}(r^{\prime })^{1/2}}{c_{0}}-1\right)\rho(r^{\prime})4 \pi r^{\prime 2}g_{rr}^{1/2}(r^{\prime})dr^{\prime}
\label{39Juan}	 
\end{equation}

where $g_{00}^{1/2}(r)/c_{0}$ is the total energy of a unit mass at $r$, including the rest
mass, while $g_{00}^{1/2}(r)/c_{0}-1$ is its gravitational energy. We have seen that $M(r)$
must be equal to the total energy (divided by $c_{0}^{2}$), which, in turn, must be equal to
the energy of the infinitly dispersed cloud minus $W$. But the former is simply the rest
energy of the cloud:

\begin{equation}
\int^{r}_{0} \rho(r^{\prime})4 \pi r^{\prime 2} g_{rr}^{1/2}(r^{\prime})dr.
\label{40Juan}
\end{equation}

Subtrating Eq. (\ref{39Juan}) from this quatity (and divigding by $c_{0}^{2})$, we have for
$M(r)$:

\begin{equation}
M(r)= \dfrac{1}{c_{0}^{2}}\int^{r}_{0}g_{00}^{1/2}(r^{\prime })\rho(r^{\prime})
4 \pi r^{\prime 2}g_{rr}^{1/2}(r^{\prime})dr^{\prime}.
\label{41Juan}
\end{equation}

It must be noted that $g_{00}(r^{\prime})$,$g_{rr}(r^{\prime})$ in this expression are not
the actual ones within the cloud but the ones that will exist after all layers above
$r^{\prime}$ had been removed, that is, the ones corresponding to the vacuum case solution
with mass $M(r^{\prime})$. This is so because layer $r^{\prime}$ is carried to infinity not
through the actual field but through the field generated by $M(r^{\prime})$. But for the
vacuum case solution the product: $g_{00}^{1/2} g_{rr}^{1/2}/c_{0}$, is equal to one. Thus,
we must have:

\begin{equation}
M(r)=\dfrac{1}{c_{0}^{2}}\int^{r}_{0} \rho 4 \pi r^{\prime 2} dr^{\prime}
\label{42Juan}
\end{equation}

which agrees which the result obtained within the full GR\cite{Landau5}. 

It must be noted that Eq. (\ref{42Juan}) it is not the sum of the masses, which is given by
Eq. (\ref{40Juan}). It is the sum of the mass elements multiplied by their corresponding
values of $g_{00}^{1/2}(r^{\prime})$. Note also that it is not the sum of the total energies
of the mass elements, because $g_{00}$ is not the actual one within the cloud extending up
to $r$, but the vacuum solution with mass $M(r^{\prime})$.

It must be remarked that the meaning of $M(r)$ is the mass that must enter the vacuum
solution that would exist if all matter beyond $r$ was removed and that it must be equal to
the total energy within $r$. But for the latter to be true, the remaining system (the matter
within $r$) must be stable. This can only be achieved by enclosing the system in an
spherical  container with radius $r$ to withstand the pressure, $P(r)$, after the outside
material, has been removed. It is only in this circumstances that there will be a vacuum
solution beyond $r$ with $M$ given by Eq. (\ref{41Juan}). The question now is: does the
presence  of the  container affect the solution at $r$? If the presence of the container,
that is massless and arbitrarily thin, was immaterial, $g_{00}(r)$ (normalized to the center
of the cloud), $g_{rr}(r)$, must be equal to the corresponding value in the actual solution
(before removing the outer matter), because the outer matter does no affect them. But we
know that this can not be, because otherwise $g_{00}$ would not depend on pressure, in
contradiction with  Eq. (11).

One may discard  the possibility of  $g_{rr}$ depending  on the enclosed pressure-volume by
the following consideration: take an arbitrarily rigid spherically symmetric solid. It is
possible to redistribute the tension (remaining in equilibrium) increasing it in the outer
parts and diminishing it in the inner ones spending an arbitrarily small amount of energy.
This can be done, for example, by heating the outer layers, since for given thermal
properties the energy needed diminishes as the elastic modulus increases. But if  $g_{rr}$
depended on the enclosed pressure-volume, its value in the inner parts would change, because
of the outwardly transferred pressure-volume. This would imply an arbitrarily large increase
of the elastic energy not compensated by a diminishing gravitational energy (which increases
slightly due to the small expansion of the body).This is in contradiction with energy
conservation, although the reader may also note another inconsistency hidden in this
argument that is independent of energy conservation.

Consequently, $g_{rr}$ must be equal at both sides of the arbitrarily thin container and,
therefore, must be given by Eq. (\ref{Juan20}), that , consistently, tell us that $g_{rr}$ 
depends only on the enclosed mass. On the other hand, for $g_{00}$   Eq. (\ref{13Juan})
shows the relevance of the enclosed pressure-volume.

From  Eq. (\ref{13Juan}) we have:
\begin{equation}
\vec{\nabla}g_{00}^{1/2}(2)-\vec{\nabla}g_{00}^{1/2}(1)=\dfrac{4 \pi G}{c_{0}^{4}}\dfrac{g_{00}^{1/2}(r) \int 3 \bar{P}dV}{r^{2}} \vec{e}_{r}
\label{42}
\end{equation}

here 2, 1 distinguish respectively the solution just  outside and just inside the container
and where the volume integral is over the walls of the container. $\bar{P}$ stand for  the
pressure in this wall (in fact a tension) no to be confused with the pressure, $P(r)$,
positive for usual fluids, which is exerted by the fluid upon the wall. Simple equilibrium 
considerations lead to the following  relationship  between the tension, $T$ on surface
elements whose vectors are within the plane tangent to the container walls and $P$.

\begin{equation*}
T= \dfrac{r P(r)}{2\Delta r }; \quad 3\bar{P}=\dfrac{rP(r)}{\Delta r}
\end{equation*}

where $\Delta r$  is the physical thickness of the container walls and where the last
equality follows from the fact  that $\bar{P}$  is the average of  the pressure in 3
orthogonal directions, and in the direction perpendicular to the walls the tension is zero.
Inserting this in  Eq. (\ref{42}) we find:

\begin{equation*}
\dfrac{1}{g_{rr}^{1/2}} \dfrac{\partial g_{00}^{1/2}(1)}{\partial r}=\dfrac{1}{g_{rr}^{1/2}} \dfrac{\partial g_{00}^{1/2}(2)}{\partial r}+ \dfrac{4 \pi G}{c_{0}^{4}}g_{00}^{1/2} r P(r).
\end{equation*}

For $g_{00}$ itself it is no necessary to distinguish between the inner and the outer value,
because it is continuous. It is  convenient to write this equation in the form:

\begin{equation}
\dfrac{1}{2g_{rr}^{1/2}}\dfrac{\partial \ln g_{00}}{\partial r}=\dfrac{1}{2g_{rr}^{1/2}}\dfrac{\partial \ln \bar{g}_{00}}{\partial r}+ \dfrac{4 \pi G }{c^{4}_{0}}g_{00}^{1/2}r P(r)
\label{43}
\end{equation}

which gives the relationship between the inner (actual solution) field strength and the
outer one given by the vacuum solution, $\bar{g}_{00}$

\begin{equation*}
\bar{{g}}_{00}(r)=c_{0}^{2} \left( 1 - \dfrac{2 GM(r)}{c_{0}^{2}r} \right).
\end{equation*}

It must be noticed that in derivating $\bar{g}_{00}$ in Eq. (\ref{43} ) $M(r)$ mus be hold
fixed, because there is no matter beyond $r$ in the hypothetical situation that  we are
considering. However, on integrating this expression to obtain the solution for the actual
distribution, $M(r)$ is not a constant. We then have for $g_{00}$:

\begin{equation}
\ln g_{00}=- \dfrac{1}{c_{0}^{2}}\int_{r}^{\infty} \left(\dfrac{2GM(r^{\prime})}{r^{\prime 2}}  + \dfrac{8 \pi G r^{\prime}}{c_{0}^{2}}P(r^{\prime}) \right) \left( 1- \dfrac{2GM(r^{\prime})}{c_{0}^{2}r^{\prime}} \right)^{-1}dr^{\prime}
\end{equation}

with the normalization $g_{00}(\infty)=c_{0}^{2}$. This is equivalent  to the expression
derived by Weinberg (REF: Weinberg pag 302) with the standard  procedure. In this procedure
the equation of hydrostatic equilibrium  is hidden  amongs the three independent equations.
Here we can derive that equation from direct considerations, obtaining:

\begin{equation}
\dfrac{1}{\sqrt{g_{00}}}\dfrac{\partial P\sqrt{g_{00}}}{\partial r}= - \rho \dfrac{\partial \ln g_{00}^{1/2}}{\partial r}
\end{equation}

where the presence of the factor $\rho$ is due to the fact that the weight of a volume
element in a gravitational field is proportional to  $\rho$. The presence of $g_{00}$ in the
left hand side comes from the fact that the total energy (including gravitational energy)
transferred by a unit force  when the body upon which is exerted is displaced by a unit of
length it is not one unit of energy, but $g_{00}^{1/2}/c_0$.

Using Eq. (43) in Eq. (45) and if $P$ is a unique function of a second order differential
equation for $M(r)$ is obtain. Also, for a politropic model ($\rho=AP^{\alpha}$, $\alpha < 1
$), we can integrate Eq. (45) to obtain a relationship between $P$ or $\rho$ and $g_{00}$:

\begin{equation}
P= \left(A (g_{00}^{\frac{\alpha-1}{2}}-1)\right)^{\dfrac{1}{1-\alpha}}.
\end{equation}

It is clear, however, that these models can not correspond exactly to finite mass object:
for $\alpha < 3/4$ as well as for the case $\alpha=1$, wich must be treated separately.

\section{Summary and Conclusions}

We have reviewed the arguments given by Einstein in the derivation of his first , non
Riemmanian, theory of gravitation (1912). We have weighed carefully these argument and
strengthened some points that could look fragile in the original presentation. We concluded
that, except for the argument proving that only matter gravitates, these arguments are
cogent enough as to leave no serious doubt about the validity of Einstein first field
equation, at least for negligible pressure, and that its ultimate rejection by Einstein was
due to a miss-identification of the culprit of the contradiction he was finding. Once we
concluded that Einstein first gravity field equation has to be correct (almost), we
proceeded to demonstrate that this equation follows  from the field equations of GR when
pressure is negligible. In general we find that the source of ``the gravitational field''
($g_{00}$) is mass density plus three times the pressure, rather than mass density alone. We
have shown that with Einstein first field equation (Eq. (\ref{6Juan})) and a basic argument
relating $g_{rr}$ and $g_{00}$ the spherically symmetric vacuum case (called Schwarzschild
solution when treated within the full GR) can be obtained immediately both with and without
cosmological constant, and have pointed out that the metric for a charged mass ``point'' can
easely be obtained in the same manner. We noted that in the static case the field  equation
remains formally equal to that in the Newtonian case (Laplacian of the field equal to zero),
only that replacing the usual gravitational potential, $\phi$, by $c$:

\begin{equation*}
c \equiv c_{0} e^{\phi/c_{0}^{2}}.
\end{equation*}

But we also noted that there is an important difference between both equations hidden in the
Laplacian, because while in the Newtonian case, being space flat, $g_{rr}$ is equal to one,
in the relativistic one $g_{rr}$ depends on $g_{00}$. It have been shown that, although in
the case of a general spherically symmetric distribution no algebraic relationship between
$g_{00}$ and $g_{rr}$ exist, still it may be solved on basic principles. To this end we
used results obtained in the vacuum case, an expression for the mass enclosed within radii
$r$ obtained from first principles and the assumption that $g_{rr}(r)$ only depends on
matter within $r$. We have written particle trajectories in a form that can easily be
compared with its Newtonian counterpart, in order to be able to see in a conceptual manner
(i.e. not merely studying the differences at various orders in $v$ and $\phi$) the origins
of the differences of particles trajectories equations around a ``point'' mass in Newtonian
and Einstenian theories. We showed that the differences are three fold. First it is the fact
that the gravitational field strength  is different from the Newtonian one, a difference
whose origin has been explained before. However, it enter in the motion equation in such a
manner that the corresponding term agrees exactly with it Newtonian counterpart. Then there
is a dependence of the gravitational ``force'' on velocity that followes immediately from
the EP (non-Riemmanian), with a term proportional to $v^{2}$. This dependence is obviously
contained in Einstein's 1912 theory. Finally it is the fact that the geometry of a plane
containing the central point it is not Euclidean. We have shown that this effect can be
reduced to replacing some simple relationships valid in the Euclidean plane by the
corresponding ones in the actual plane, that can be derived from elementary geometrical
considerations. Within this analysis we have computed the angular velocity of the perihelion
of quasi-Keplerian orbits and found that $2/3$ of it are due to the velocity dependence
(i.e. to EP), the other third being due to non-Euclideanity. We have also used the same
approach for computing the deflection of light rays, showing in a simple manner the well
known result that $1/2$ of the effect is ``Newtonian'' and the other half comes from
non-Euclidianity, with the velocity dependent term playing no role.

By considering the gravitational interaction of two ``point'' masses, we have made patent
how the assumption of flat space causes the non-conservation of momentum that plagued
Einstein first gravitational theory. We have then discussed a possible path of  development
of GR that could have happened if, after realizing that the flatness of space was untenable,
Einstein had  removed the error from his first gravitational theory and continued his
initial research program.

We have discussed Einstein arguments showing that pressure can not gravitate and why it
fails. Then we reviewed the argumental line leading to the presence of pressure in Eq.
(\ref{13Juan}), showing that it is a direct consequence of ``general covariance'' and,
therefore, closely related to energy-momentum conservation. We have concluded that of all
question treated in this work this is, arguably, the most difficult to have been anticipated
in 1912 and, although we have thought of an argument proving this fact, we have not
presented it here because it involves non-static fields. Finally, we have used Einstein
argument relating gravitational mass and total energy to obtain an expression for the mass
enclosed within radii $r$ in a general  spherically symmetric distribution, which is
instrumental in deriving the metric in this case.This we have done in two manner, first
using the expression for $g_{rr}$ obtained with the mentioned argument in Eq. (11) and then
by extending this argument to obtain $g_{00}$ through direct physical considerations. We
also give a direct derivation of the equation of hydrostatic equilibrium and discuss how to
use it together with the expression for $g_{00}$ and $P(\rho)$ to obtain the equilibrium
configuration of a self-gravitating non-rotating cloud. In the standard procedure, using
Einstein equations (GR), this is obtained in a non very transparent manner.

The general aim of this work has been to try to understand the most elementary result of GR
in terms of basics principles. To this end Einstein's 1912 gravity equation has been
instrumental along with the EP, energy-momentum conservation and the mass energy
relationship. The results that we have derived are not only of heuristic value, which has
been our main goal, but are amongst the most relevant result of GR for astrophysics.

In all other fields of physics we are used to dealing with trivial cases in a simple manner,
identifying the relevant basic principles in a problem to anticipate the aspect of it that
can quickly be learned and combining elements of different solutions (not necessarily
superposing them) to gain some insight into the actual problem. In GR, however, this usual
heuristic is almost entirely lacking. Confronted with a particular problem it is the
geometrical symmetry that determine its treatment and then we switch on our elegant
mathematical formalism to obtain the solution, gaining little insight in the process. Then,
when extracting the physical meaning of the solution, it is not done, in our opinion, in the
most enlightening manner, even in the simple cases in which more clarifying procedures seems
possible.

Take the case of the spherically symmetric vacuum solution. In the homologous case in
electromagnetism we know in advance that there is just one quantity, the electrostatic
potential, to be determined. How could a point endowed with just one type of charge
determine two different (not a function of each other) radial dependences?. In the present
case, however, we have two equations for $g_{rr}$ and $g_{00}$ and only after integrating
them it is found, in an nontransparent manner, that they are simply related, a relationship
that could has been advanced based on the EP.

In the case of a spherical cloud in equilibrium we have, in GR, three independent equations.
Certain combination of them must give the hydrostatic equilibrium equation, but this is not
clear at all in most expositions.

For the spherically symmetric vacuum solution particle trajectories can be determined
exactly in the most elegant manner using Hamilton-Jacobi formalism. But then the origin of
the relativistic effects is completely masked. Approximate treatments as that given by
Einstein\cite{A.E.libro} are somewhat more transparent, but still, in our opinion,
insatisfactory. General three-dimensional treatments\cite{Landau4} are much more helpful,
but still, having to compute explicitly the three-dimensional affine connections seems
without proportions with the intrinsic simplicity of the problem. Here we have shown how
simply can this problem be treated exactly, and not for the reason of the economy of work
but to gain in transparency and insight.

The questions we have treated here are not specialized ones, but very basic and of interest
for any physicist with an eye on GR. The fact that at least some of them has not been
treated, or rather, that they are not treated where obviously they should have been, seems
to us a sort of anomaly, since there is no question here that could not has been treated and
be widely known at least 80 years ago.

It is true that in GR we are deprived of our most valued source of intuition, the
``solidity'' of space, that time is not what it used to be and then we have the
non-linearities, tensorial field and source, etc. But this should be an stimulus to work
harder and not something to make us stagger back and stick to an infallible but obscure
formalism renouncing to develop a detailed physical intuition of GR for fears of being
betrayed by it too often.

The terrain is treacherous, and we may atest to it. The simple and, we hope, very easy to
follow derivations given here were (some of them) not easy to obtain. When trying to reduce
the result searched for to a combination of simple ideas, the above mentioned complexities
of GR made progress difficult, but now that we have managed to treat the questions presented
here in a consistent manner, we can only hope that this questions enhance the understanding
of GR of the reader as much as it did with ours.

In the presentations of GR the dependence of the relative rate of two clocks on a
gravitational field on the diffference of potential between them and the related
gravitational redshift are directly derived in a simple manner from the the EP . Some of
them even discuss the non- Euclidianity of space in a rotating system in flat time-space.
But beyond this point direct derivations from basic principles are no longer given , not
even heuristics ones, being replaced by mathematical considerations. This mimic the
development of GR : when Einstein met with overwhelming difficulties in 1912 , he switched
to a mathematically guided line of research. In consequence, elementary facts and
considerations, as how do genuine gravitational static gravitational fields affect the space
geometry,  how and why the passive and active couplings of matter to gravity (the field
$g_{00}$) do differ and why the latter include pressure, or how the implications of the EP
in the case of flat space combine with the effect of the spatial curvature in determining
particle trajectories, are not discussed. It is the aim of this work to contribute to ease
this situation, that would  have been accepted in no other field of physics.

We are aware of the fact that , although some of the simple derivations given here could be
used as a base for an elementary exposition of GR , such as it stands it is rather an
advanced one , being addressed to those who know well the standard presentations of the
facts treated here. Our goal has been to make patent to those readers the concepts and
principles involved in these facts and how simply could they be derived. We also want to
remark that, even when we make some comments on the history of GR , this is not even
partially a work on the history of physics. This is a work on physics, and those comments
have been made to stress underlying conceptual issues. Nevertheless, we think that the
historical facts that we comment are basically correct, although when we refer to Einsteins
conceptions with respect to some question at a given date it may be only our best
interpretation of what we actually know.

\end{document}